\documentclass[%
 reprint,
 amsmath,amssymb,
 aps,
]{revtex4-2}
\renewcommand{\selectlanguage}[1]{}
\usepackage{graphicx}
\usepackage{dcolumn}
\usepackage{bm}
\usepackage{tikz}
\usetikzlibrary{shapes.geometric, arrows.meta, positioning, calc, shapes.symbols, shapes.misc, decorations.markings, shadows}
\usepackage{amssymb}  
\usepackage[ruled,vlined,linesnumbered,algosection,algoruled,commentsnumbered]{algorithm2e}
\SetKwBlock{Try}{try}{end}
\SetKwBlock{Catch}{catch}{end}
\usepackage{booktabs}
\usepackage{placeins}
\usepackage{soul}
\usepackage{multirow}
\usepackage{verbatim}

\usepackage{hyperref}

\begin{document}

\preprint{APS/123-QED}

\title{Can Large Language Models Correctly Interpret Equations with Errors?}

\author{Lachlan McGinness}
 \email{Lachlan.McGinness@anu.edu.au}
\author{Peter Baumgartner}%
\affiliation{%
 Australian National University \\
Commonwealth Scientific and Industrial Research Organisation
}%


\begin{abstract}
This paper explores the potential of Large Language Models to accurately extract and translate equations from typed student responses into a standard format. 
This is a useful task as standardized equations can be graded reliably using a Computer Algebra System or a Satisfiability Modulo Theories solver.
Therefore physics instructors interested in automated grading would not need to rely on the mathematical reasoning capabilities of Language Models.  

We used two novel frameworks to improve the translations. The first is consensus where a pair of models verify the correctness of the translations. The second is a neuro-symbolic LLM-modulo approach were models receive feedback from an automated reasoning tool.
We performed experiments using responses to the Australian Physics Olympaid exam.
We report on results, finding that no open-source model was able to translate the student responses at the desired level of accuracy. 
Future work could involve breaking the task into smaller components before parsing to improve performance, or generalizing the experiments to translate hand-written responses.
\end{abstract}

\maketitle


\section{\label{sec:Introduction}Introduction}

The Australian Physics Olympiad is an annual physics exam in Australia with approximately fifteen hundred 16-18 year-old participants. For the last five years the exam was conducted online with a combination of question types, where students are allowed to submit their responses either in a typed format or by scanning hand-written working. Students sit the test in early August and it is graded by a team of approximately fifteen markers in a 48-hour period two weeks later. 

\begin{figure*}
\centering
    \scalebox{0.84}{
    \begin{tikzpicture}[
        scale=0.9,
        every node/.style={transform shape},
        note/.style={
            draw=gray!30,
            fill=gray!5,
            rounded corners=5pt,
            text width=3cm,
            minimum height=2cm,
            align=center,
            font=\small
        }
    ]
    \tikzset{
        thought bubble/.style={
            rectangle,
            draw,
            rounded corners=10pt,  
            fill=blue!10,
            text width=3cm,       
            minimum width=2cm,    
            minimum height=1.2cm, 
            align=center,
            font=\footnotesize    
        },
        processor/.style={
            cylinder,
            cylinder uses custom fill,
            cylinder body fill=orange!30,
            cylinder end fill=orange!10,
            shape border rotate=90,
            aspect=0.5,
            minimum width=2.5cm,
            minimum height=3cm,
            align=center
        },
        code block/.style={
            rectangle,
            draw=gray!50,
            fill=black!5,
            rounded corners=3pt,
            minimum width=3cm,
            minimum height=2cm,
            align=left,
            font=\ttfamily\small
        },
        score display/.style={
            circle,
            draw=green!50!black,
            fill=green!20,
            minimum size=2cm,
            font=\Large\bfseries
        }
    }
    
    \node[thought bubble] (input) at (0,0) {
        \textbf{Student Input}\\
        My final velocity\\
        is given by:\\
        $v^2 = u^2 + 2as$\\
        this is because...
    };
    
    \node[processor] (llm) at (3.5,0) {
        \textbf{LLM}\\
        \small{Natural Language}\\
        \small{Parser}
    };
    
    \node[code block] (parsed) at (7,0) {
        [Eq(v, u),\\
         Eq(v, w**2/(2*S))]\\
    };
    
    \node[processor] (engine) at (10.5,0) {
        \textbf{Reasoning}\\
        \textbf{Engine}\\
        \small{Symbolic Analysis}
    };
    
    \node[score display] (score) at (13.5,0) {
        3/3
    };
    
    \foreach \i/\j in {input/llm, llm/parsed, parsed/engine, engine/score} {
        \draw[
            -stealth,
            line width=2pt,
            red!50!blue,
            decoration={
                markings,
                mark=at position .5 with {\arrow{stealth}}
            },
            postaction={decorate}
        ] (\i) -- (\j);
    }

    \draw[blue, dashed, thick] 
    ($(llm.north west)+(-0.2,0.7)$) rectangle 
    ($(engine.south east)+(0.6,-0.6)$);

\draw[red, dashed, thick] 
    ($(llm.north west)+(-0.1,0.6)$) rectangle 
    ($(engine.south east)+(-6.3,-0.5)$);

    \node[blue, font=\bfseries] at ($(parsed.south)+(0,-1)$) {AlphaPhysics};
    \node[red, font=\bfseries] at ($(parsed.south)+(-3.3,-1)$) {Focus of this paper};
    \node[align=center, font=\bfseries] at ($(parsed.south)+(0,0.4)$) (caption) {Symbolic\\ Representation};
    \end{tikzpicture}
    }
    \caption{\label{fig:alphaphysics-pipeline}The AlphaPhysics Pipeline. The student response is parsed by an LLM. A symbolic reasoning engine then determines the student's grade.}
\end{figure*}
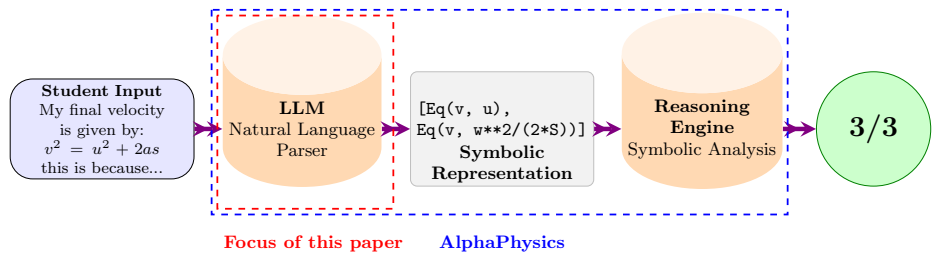

In this paper we propose a two-step pipeline for the automated grading of physics exams that we call AlphaPhysics, see Figure \ref{fig:alphaphysics-pipeline}. Although there is no official connection to Google Deepmind, AlphaPhysics follows the same structure as AlphaGeometry where a neural model guides a symbolic deduction engine to solve Olympiad geometry problems \cite{AlphaGeometry}. 

The focus of this paper is the first step, which calls for Large Language Models (LLMs) to extract and translate equations from student responses into a standardized format. The second step would use an automated reasoning tool such as a Satisfiability Modulo Theories (SMT) solver or Computer Algebra System (CAS) to verify whether the translated equations are equivalent to those required in the solution. 
This means that instructors interested in automated grading would not need to rely on the notorious, but improving, mathematical reasoning of LLMs \cite{Cobbe2021Training,Hendrycks2021Measuring,Ahn2024Large,Zhang2024Careful} but instead know that reasoning is being performed by reliable systems with guarantees\cite{Barbosa2023Challenges,Guidotti2023Leveraging,Winterer2024Validating}. Knowing that a grading system has reliable reasoning over algebraic formulas may influence a physics instructor's choice to adopt automated grading.

SMT solvers and CASs require inputs to adhere to strict syntax requirements \cite{SMTLIB,Lachnitt2024IsaRare}. This makes these tools incapable of interfacing with student responses on their own. LLMs are theoretically capable of reading student responses and translating to the required syntax. This paper evaluates LLM performance on this task. 

We systematically investigate the performance of open-source LLMs of different sizes to preprocess a sample of 200 typed responses to a physics question which requires a combination of interwoven words and equations. 
We evaluate two techniques for improving the LLM translations, including inter-LLM consensus and the LLM-Modulo framework where the LLM receives feedback from a symbolic reasoning engine \cite{Subbarao2024LLMmodulo}. We establish a relationship between model size, computational expense and accuracy in translation. We report on the main types of errors made by LLMs and discuss the computational cost of running them. 

\section{Background\label{sec:Background}}

\subsection{Automated Grading for Physics}

Automated grading is a long-standing area of research, with one of the first prominent topics being Automated Short Answer Grading (ASAG) \cite{Burrows2015Era}. In the early 2000s the main approach to ASAG was to use regular expressions  to match key words or phrases to those in the marking scheme \cite{Hasanah2016Review}. This was not scalable as it required instructors to think of all ways that students could express the correct answer. 

Starting in 2008, there was a significant paradigm shift to using automated feature extraction and machine learning techniques for grading  \cite{Mat2018Text,Wang2008Assessing,Ziai2012Short}. These approaches were more robust, but required large amounts of marked examples as training data. Some recent machine learning methods have aimed to reduce the amount of data required for grading and make the computer-generated grades more explainable \cite{McGinness2024CONFOLD}.

Recently, generative pre-trained transformers, also known as Large Language Models (LLMs), were used to automatically grade  short answer questions \cite{Grevisse2024LLM-based, Tobler2023Smart, Yan2024Exploring, Kortemeyer2024Automated}. Domain-specific pre-training and fine-tuning improve the grading accuracy of LLMs \cite{Bonthu2024Framework, Menezes2024AI-Grading}. 
Studies have explored LLMs' capabilities to grade using rubrics \cite{Senanayake2024Rubric, Yan2024Exploring} and found that for complex questions, LLMs struggle to recognize merit in diverse responses.  Another study found that specific adjectives and adverbs cause LLMs to falsely assume that the answers are correct \cite{Filighera2024Cheating}. 

Physics problems are very difficult to mark automatically because they require a wide range of inputs including numerical responses (often with units and directions), short answers, algebraic expressions, diagrams and plots of data/functions. However, the recent development of multi-modal LLMs mean that they now show some promise as a feature extraction tool for the automated grading of plots and diagrams \cite{Kortemeyer2024Automated}.

Previous studies which have used LLMs to grade physics exams \cite{Kortemeyer2024Automated, Mok2024Using, Chen2024Achieving, Chen2025Grading, Kortemeyer2025Assessing} found that LLMs may hallucinate. Furthermore there are no guarantees that LLM mathematical reasoning is correct. There are also a number of ethical considerations in the Australian Government's AI Ethics Principles \cite{AustraliaAIEthics} which have not explicitly been considered in previous studies including environmental well-being, transparency and privacy protection.

\subsection{Ethical Dimensions of Grading}
\paragraph{Environmental Impact} Outside of the grading literature, studies have criticized the use of LLMs for their large large electrical cost and  environmental impact \cite{Bender2021Dangers}. 
Although most LLMs use considerable amounts of computational power (and therefore electricity), to our knowledge there are no studies which account for or attempt to reduce the computational cost of grading physics exams with LLMs.

\paragraph{Tranparency} Existing studies \cite{Kortemeyer2024Automated, Mok2024Using, Chen2024Achieving, Chen2025Grading, Kortemeyer2025Assessing} which use Large Language Models to grade physics exams used proprietary models. 
These models and their system prompts are changed regularly by the companies which own them. This leads to variations in the performance of grading systems over time. Neither instructors nor students are able to access the models, know their architecture or know when they are updated.
Unless using legacy versions of models, instructors cannot even be sure that all students within a single grading batch have been graded by the same model.

This issue of transparency can be addressed by using open-source models. If the model is open-source then both the instructor and the students are able to know the parameters of the model(s) being used. We note that it would also be possible to run open-source models on GPU servers to improve their power efficiency, although we are not proposing this universally because of privacy.

\paragraph{Privacy} To our knowledge, studies have only explored the use of frontier (state-of-the-art), proprietary LLMs for grading physics exams. In these cases student responses must be uploaded to an external server to be graded. Some privacy concerns can be alleviated through cloud-based services which give guarantees that this information will only be stored in certain countries and not used for training future models. 
Unfortunately this may not satisfy the policies of all educational institutions. 

This study addresses the existing gap of evaluating LLMs which are small enough to run on consumer-grade hardware (a graphics card or GPU that can be found on a `normal computer'). 
The ability to grade on their own computer with easy-to-use tools like Ollama \cite{Ollama} and can make LLM-based methods available to more instructors. Smaller models also have the benefit of performing less computations, therefore using less power. 
Running models locally also makes instructors aware of the compute that they are using, exposing the normally hidden electricity usage and environmental cost.

\paragraph{Contestability and Reliability}
The AlphaPhysics pipeline uses LLMs only for pattern recognition; to extract equations. These equations can be easily seen, contested and corrected by a student or teacher. 
A rigorous source of reasoning over algebraic formulas, such as Automated Theorem Provers, can then be used to ensure reliable grading.

\subsection{Automated Reasoning}


Satisfiability Modulo Theories (SMT) solvers are a class of Automated Theorem Prover that handle mathematical reasoning. 
They generally deal with formulas containing mathematical operators over integers and real numbers.
For a more detailed introduction to SMT solvers we recommend \cite{Moura2011SMT}. 
We use Z3, an SMT solver \cite{Z3} to provide an LLM with feedback on its translated equations. Z3 is an open-source SMT solver which consistently performs well in the SMT-LIB competition \cite{Beyer2025SMTCOMP}. 

Computer Algebra Systems (CASs) are designed to handle a wider variety of tasks than SMT solvers. 
They can aid users by simplifying expressions or performing algebraic preprocessing. For more information on the fundamentals of CAS we point the reader to \cite{Davenport1993CAS}.  
We use a python-based CAS called SymPy  \cite{Sympy}. 
SymPy is an open-source python library for symbolic manipulation involving algebra, calculus, discrete mathematics and matrices \cite{Meurer2017Sympy}.

Unfortunately, neither SMT solvers nor CASs are able to interpret natural language expressions. Instead they require preprocessing to bring the students' equations into a standardized symbolic representation as shown in Figure \ref{fig:alphaphysics-pipeline}. If LLM translation is accurate enough we envisage that a combination of an SMT solver and a CAS would be used to rigorously check if the translated student equations match those in the marking scheme.

\section{Methods}

\subsection{The Physics Olympiad Question}

We chose question four from the 2023 Australian Physics Olympiad for grading because it required the students to give an algebraic response and student working often included many interwoven lines of algebra and free-form text. This made the task of extracting the equations more challenging for the models. 

The full question can only be understood in the context of the preceding questions which are provided (with the accompanying diagrams) in Appendix \ref{app:question}. An abbreviated version of the question is as follows:

\begin{quote}
    \textit{A group of students walk with constant speed, $u$. They travel a distance, $S$, in a straight line then return. Mali, walks the two-way journey on a travellator which moves parallel to the group's original velocity at speed $w$. Mali walks at a constant speed, $v$ (relative to the ground beneath them), for the entire journey. Determine speed $v$ such that Mali and the group return at the same time. }
\end{quote}

The question also asks students to explore limiting behaviors of their equations. Note that in this study we were not yet interested in whether student answers corresponded to the correct solution. We investigate whether the list of equations that the LLM found in the response matches the equations that the markers thought were present in the student responses. 

A team of four graders graded every student response to this question.
They discussed ambiguous cases to decide the final score for each response. 
These two hundred cases were then independently checked by the Australian Physics Olympiad Director who agreed with the allocated scores and the list of equations included in each response. 

\subsection{Prompting Techniques}

In our experiments we investigated frameworks which could improve the accuracy of an LLM without increasing its size:
\begin{itemize}
    \item \textbf{Direct Method} - method used in previous studies \cite{Kortemeyer2023Toward,Chen2024Achieving,Mok2024Using,Kortemeyer2024Grading,Kortemeyer2025Assessing} where a model is given a prompt to extract equations from a student response. Note that the Direct Method may include significant prompt engineering.
    \item \textbf{LLM-Modulo SymPy} - after the Direct Method is applied, the model is warned if its response does not match the required syntax. The model is then given a chance to improve its response. See Algorithm \ref{alg:llm-modulo}.
    \item \textbf{LLM-Modulo Z3} - similar to the LLM-Modulo SymPy approach, but the model is given specific feedback for the type of syntactic error. For example ``symbol `U' is undefined''.
    \item \textbf{Consensus} - builds on the LLM-Modulo SymPy procedure by comparing two models' responses using Z3. If they are not equivalent then the models are given three attempts to reach consensus, see Algorithm \ref{alg:consensus}. If consensus is not reached then the model keeps its own response. 
\end{itemize}
\noindent The sample of 200 responses contained 30 blank responses which were parsed automatically without calling an LLM. Each technique was tested with a range of models, as described in Section \ref{sec:models}, using the metrics described in Section \ref{sec:metrics}.

Algorithm \ref{alg:llm-modulo}, describes the exact LLM Modulo feedback loop. 
In summary, after the LLM has extracted equations from the student answer, each equation is parsed by SymPy. 
If no syntax errors occur then this is accepted as the final list of equations. 
If there is a syntax error, then the model is given its list of equations and the associated errors and prompted to repair the list of equations. 
This produces an updated list of equations. If none of the updated equations contain syntax errors the new list is accepted. 
Otherwise the model is prompted to repair the updated list. 
In our experiments, the LLM was given three chances to repair the list of equations before we accepted the list with the syntax errors.

\begin{algorithm}[h!]
\scriptsize
\caption{LLM\_Modulo Algorithm: gives LLM feedback}
\label{alg:llm-modulo}
\KwIn{$student\_answer$, $max\_attempts$, $parser$}
\KwOut{$modulo\_list$ \tcp*{parsed equations or final best attempt}}

\SetKwFunction{FParse}{Attempt\_parse}
\SetKwFunction{FGetRaw}{Get\_llm\_direct\_response}
\SetKwFunction{FGetRepair}{Get\_llm\_repair\_response}

$direct\_list \gets$ \FGetRaw{$student\_answer$}\;
$correct\_syntax, error\_messages \gets$ \FParse{$direct\_list$, $parser$}\;

\If{$correct\_syntax = True$ }{ 
    $modulo\_list \gets direct\_list$ \tcp*{All equations syntactically correct.}
    \Return{$modulo\_list$}\;
}

$attempts \gets 0$\;
$modulo\_list \gets direct\_list$\;

\While{$attempts \leq max\_attempts$}{
    $attempts \gets attempts + 1$\;
    $modulo\_list \gets$ \FGetRepair{$student\_answer$, $modulo\_list$, $error\_messages$}\;
    
    $correct\_syntax, new\_error\_messages \gets$ \FParse{$modulo\_list$}\;
    
    \If{$correct\_syntax = True$}{
        \textbf{break}\;
    }
    
    $error\_messages \gets error\_messages \cup new\_error\_messages$\;
}

\Return{$modulo\_list$}\;
\end{algorithm}

Algorithm \ref{alg:consensus}, describes the exact consensus algorithm. It begins with two different models (we chose two models of approximately the same size) each undertaking the LLM\_Modulo algorithm, receiving feedback from Z3. 
Z3 then checks the equivalence of the lists of equations from the two models using the procedure described in the next paragraph. 
If they are equivalent then consensus is reached and the list of equations accepted. 
If they are not equivalent then each model is provided both lists of equations and the original student response and asked to produce a new list using the LLM modulo procedure. 
The updated lists are compared by Z3. The models are given three attempts to reach consensus, otherwise each model keeps its own response. 
In the case where a user only wanted one final answer, they could randomly choose one of the two models' responses or preferentially choose one model's responses over the other's.

\begin{algorithm}[h!]
\scriptsize
\caption{Consensus\_Algorithm: LLM Equation Agreement}
\label{alg:consensus}

\KwIn{$Model1$, $Model2$, $student\_response$, $max\_attempts$, $parser$}
\KwOut{$Model1\_equations$, $Model2\_equations$}

\SetKwFunction{FModulo}{LLM\_Modulo}
\SetKwFunction{FCheck}{Check\_equivalence}
\SetKwFunction{FConsensus}{Create\_consensus\_prompt}
\SetKwFunction{FGetResponse}{Get\_LLM\_response}

$Model1\_equations \gets$ \FModulo{$student\_response$, $parser$, $Model1$}\;
$Model2\_equations \gets$ \FModulo{$student\_response$, $parser$, $Model2$}\;

$attempts \gets 0$\;

\While{$attempts \leq max\_attempts$}{
    $attempts \gets attempts + 1$\;
    $are\_equivalent \gets$ \FCheck{$Model1\_equations$, $Model2\_equations$}\;

    \If{$are\_equivalent = True$}{
        \Return{$Model1\_equations, Model2\_equations$}\;
    }
    
    $consensus\_prompt \gets$ \FConsensus{$Model1\_equations$, $Model2\_equations$, $student\_response$}\;
    
    $Model1\_equations \gets$ \FGetResponse{$consensus\_prompt$, $Model1$}\;
    $Model2\_equations \gets$ \FGetResponse{$consensus\_prompt$, $Model2$}\;
}

\Return{$Model1\_equations, Model2\_equations$}\;
\end{algorithm}

Additional algorithms in Appendix \ref{app:algorithms} are helper functions used in LLM\_Modulo and Consensus, these are briefly described in the following paragraphs. 
These functions check for equivalence and entailment as follows.
For two lists of equations $E_1$ and $E_2$, we say that $E_1$ entails $E_2$ iff for every $e_2$ in $E_2$ there is an $e_1$ in $E_1$ such that $e_1$ entails $e_2$. We say that $E_1$ is equivalent to $E_2$ if $E_1$ entails $E_2$ and $E_2$ entails $E_1$. All reasoning is carried out with respect to  real arithmetic in Z3.

As an example, consider the following two lists of equations $[v=u+w, v=u+w]$ and $[v-u=w, v=u-w]$ which we call $E_1$ and $E_2$ respectively. $E_2$ entails $E_1$ as the first formula in $E_2$ entails both formulas in $E_1$. However $E_1$ does not entail $E_2$ as $v=u-w$ is not entailed by any formula in $E_1$. Hence the two lists are not equivalent.

\begin{table*}
    \centering
    \caption{\label{tab:raw-data} Raw data for each model and each technique. In most cases models were paired with the other model of the same size for consensus, the exceptions were these pairs: Llama3.1 405B and Deepseek 671B, Llama 3.2 3B and Gemma2.2B, and Deepseek 1.5B and Llama3.2 1B. Note that the 2-3B parameter models do not include a thinking model.}
    \resizebox{\textwidth}{!}{%
    \begin{tabular}{l|*{12}{c}}
        \hline
        \textbf{Model} & 
        \multicolumn{1}{c}{Llama 3.1} & 
        \multicolumn{1}{c}{Deepseek} & 
        \multicolumn{1}{c}{Llama 3.1} & 
        \multicolumn{1}{c}{Deepseek} & 
        \multicolumn{1}{c}{Phi4} & 
        \multicolumn{1}{c}{Deepseek} & 
        \multicolumn{1}{c}{Mathstral} & 
        \multicolumn{1}{c}{Deepseek} & 
        \multicolumn{1}{c}{Llama 3.2} & 
        \multicolumn{1}{c}{Gemma} & 
        \multicolumn{1}{c}{Deepseek} & 
        \multicolumn{1}{c}{Llama 3.2} \\
        \hline
        Model Size (B) & 405 & 671 & 70 & 70 & 14 & 14 & 7 & 7 & 3 & 2.2 & 1.5 & 1 \\
        Quantisation (bit) & 16 & 16 & 16 & 16 & 16 & 16 & 16 & 16 & 16 & 16 & 16 & 16 \\
        \hline
        \multicolumn{13}{c}{\textbf{Direct}} \\
        \hline
        Time (s) & NA & NA & NA & NA & 4,460 & 90,001 & 570 & 17,141 & 209 & 183 & 2,946 & 146 \\
        Tokens & 5,705 & 105,836 & 5,257 & 109,815 & 5,582 & 117,626 & 6,625 & 132,484 & 6,740 & 6,716 & 175,361 & 10,383 \\
        Correct Answers & 154 & 169 & 149 & 166 & 141 & 129 & 80 & 110 & 75 & 53 & 80 & 33 \\
        \hline
        \multicolumn{13}{c}{\textbf{LLM-Modulo SymPy}} \\
        \hline
        Time (s) & NA & NA & NA & NA & 4,604 & 91,010 & 673 & 17,422 & 211 & 321 & 3,024 & 156 \\
        Tokens & 5,900 & 107,884 & 5,326 & 112,575 & 5,769 & 118,928 & 7,893 & 134,657 & 6,776 & 12,059 & 179,960 & 10,978 \\
        Correct Answers & 164 & 169 & 149 & 166 & 141 & 130 & 80 & 110 & 75 & 53 & 80 & 33 \\
        \hline
        After Consensus & 176 & 176 & 156 & 173 & 140 & 132 & 85 & 98 & 72 & 57 & 82 & 71 \\
        \hline
        \multicolumn{13}{c}{\textbf{LLM-Modulo Z3}} \\
        \hline
        Time (s) & NA & NA & NA & NA & 8,278 & 128454 & 885 & 22,447 & 220 & 438 & 3,508 & 188 \\
        Tokens & 7,146 & 122,038 & 6,496 & 116,469 & 7,085 & 114597 & 10,387 & 151,793 & 6,838 & 15,818 & 208,440 & 13,144 \\
        Correct Answers & 169 & 173 & 151 & 167 & 146 & 143 & 85 & 123 & 74 & 54 & 89 & 61 \\
        \hline
    \end{tabular}%
    }
\end{table*}

We use equivalence for the consensus framework and to compare the LLM-extracted lists of equations to those provided by the markers as a measurement of accuracy. 
Student working is often messy, out of order, or contains repeats; however most physics instructors do not wish to penalize students for this. Our definition for equivalence allows for this, as long as two lists of equations contain the same information, a repetition or change in order is not considered a difference.

\subsection{Selected Models}
\label{sec:models}
In this experiment we consider only open-source models, see Table \ref{tab:models} in Appendix \ref{app:models}. This ensures that the model architecture and number of parameters is well-known. This will allow us to discover the smallest model size which can accurately translate student equations from natural language text. Half of the models belong to a category called `reasoning' or `thinking' models. 
Thinking models are LLMs which are specifically trained to produce a Chain of Thought that is not intended to be shown to the user, contained within $<\backslash$think$>$ markers. 
Rather than answering a question directly, the model first gives itself `space to think' before producing its final answer tokens. 
This often improves LLM accuracy but greatly increases the computational cost. 

Our experiments mostly focus on smaller models that can be run on consumer-grade hardware. All models with less than seventy billion parameters were run on a local machine with an Intel Core-i9-13900K CPU (3-5.8GHz), 64GB of DDR5 (4800MT/s) and an NVIDIA GeForce RTX 4060Ti GPU with 16GB of dedicated GPU memory. 

\subsection{Evaluation Techniques}
\label{sec:metrics}
To evaluate each model and technique we collected each of the following pieces of information:
\begin{itemize}
    \item \textbf{Wall Time} - Wall time, is the actual real-world time that a computer spends generating the output. It is a rough measure of computational expense and therefore electrical cost, environmental impact, and practicality. Wall time was not collected for models which could not be run on local systems as it is not an accurate measure of the computational expense nor comparable with other models.
    \item \textbf{Accuracy} - LLMs were asked to provide their equations in SymPy format. This was chosen because we use SymPy to solve student equations in the next step of the pipeline \cite{Baumgartner2025AlphaPhysics}. A parser was used to reliably convert each response from SymPy format to Z3 to verify the equivalence of the LLM list of equations to the list of equations that the marking team believed was present in the student responses (see Algorithm \ref{alg:check-equivalence}). The portion of responses where the two lists were found to be equivalent was reported as the accuracy score. 
\end{itemize}

\section{Results}

\subsection{Quantitative Results}
The raw data from the original experiments are shown in Table \ref{tab:raw-data}. After the initial data were taken, additional experiments were performed with models which have lower precision weights, see Table \ref{tab:raw-data2} in Appendix \ref{app:results}. 
Note that in 3.5\% of responses, the graders could not agree exactly which equations the student intended to write. 
Therefore we would say that 3.5\% of responses are ambiguous and we could not expect any grader, human or AI, to agree with more than 96.5\% accuracy. 

\begin{figure*}
    \centering
    \includegraphics[width=1.0\textwidth]{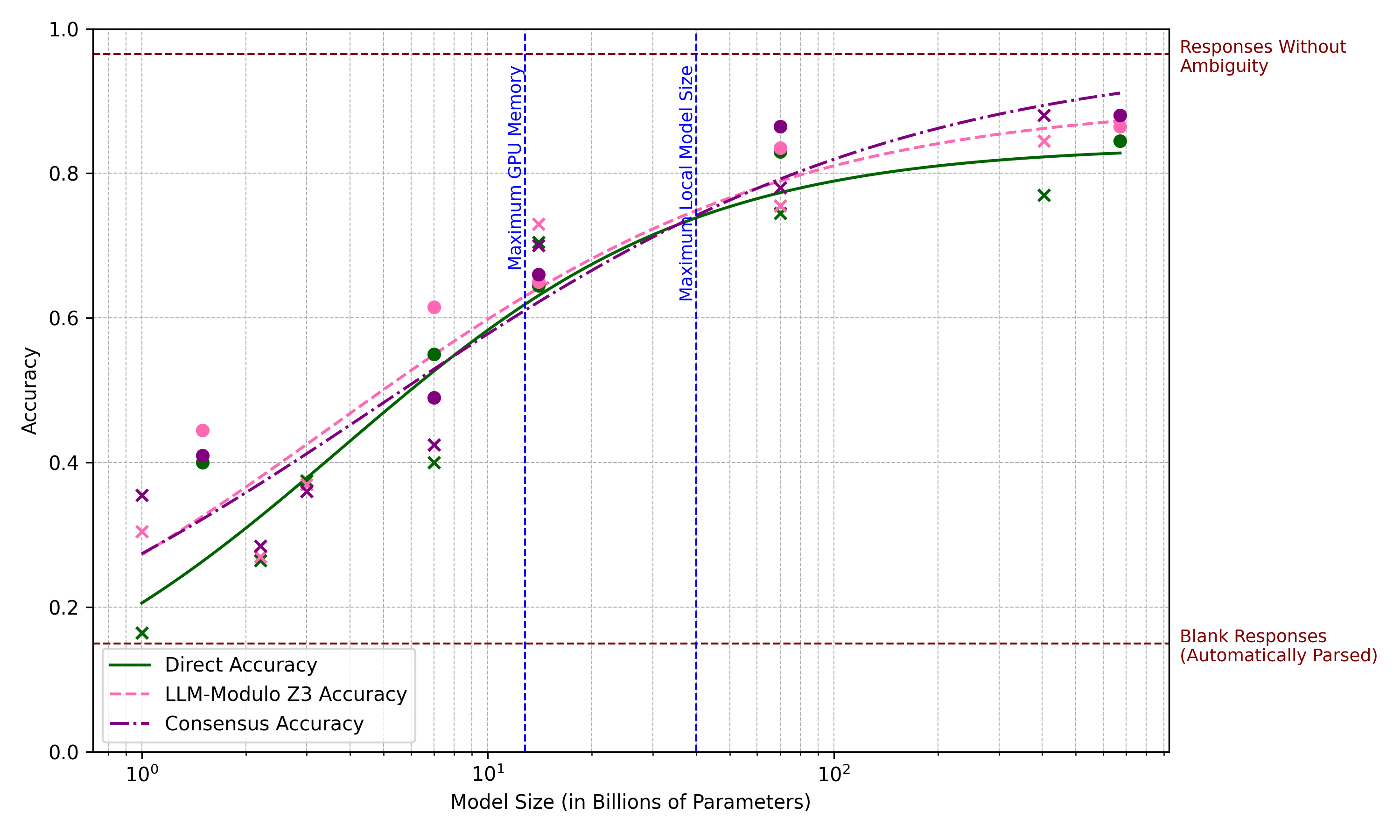}
    \caption{Plot of accuracy against model size. Colour indicates the technique being used. Circles correspond to thinking models (like Deepseek) while crosses refer to other models. Blue vertical lines indicate the maximum model size that fits in the GPU memory and the maximum model size which can be run on the hardware. Sigmoid functions show the overall trend of the data. In 3.5\% of responses the graders could not agree on what the students intended to write. Therefore upper maroon horizontal line indicates the 96.5\% accuracy where responses become ambiguous.}
    \label{fig:model_accuracy_log}
\end{figure*}

Table \ref{tab:raw-data} shows that the LLM-Modulo SymPy barely improves performance. Figure \ref{fig:model_accuracy_log} shows that as model size increases so does performance. Figure \ref{fig:model_accuracy_log} demonstrates that small models receive greatest improvement when given feedback about syntax, especially thinking models. On the other hand, larger models gain more benefit from discussion and checking their answers with others.

\begin{figure*}
    \centering
    \includegraphics[width=1.0\textwidth]{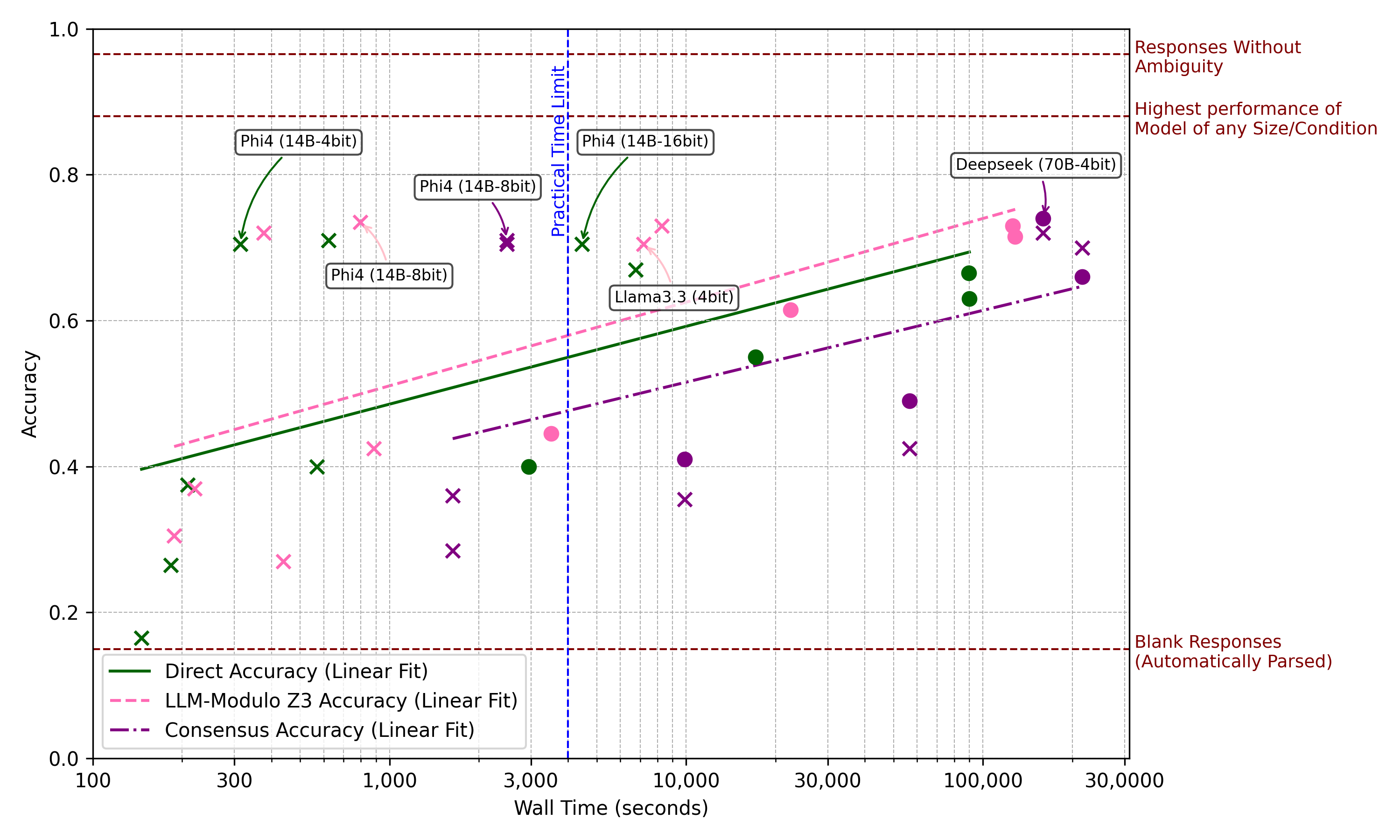}
    \caption{Model Accuracy vs. Time to Run. Colour is used to indicate the technique. Circles correspond to thinking models (DeepSeek), all other models have crosses. Three horizontal maroon lines indicate the number of blank responses, the highest performance of any combination of models, and the highest expected performance as the final 3.5\% of responses were determined by the markers to be ambiguous. }
    \label{fig:wall_time_accuracy_log}
\end{figure*}

In Figure \ref{fig:wall_time_accuracy_log} we can see that the consensus (purple) points lie to the right of their (pink and green) counterparts. 
This means that the consensus trendline is also translated to the right. This indicates that the consensus process greatly increases computational expense but only makes a small increase to overall accuracy. 

Here we provide an order of magnitude estimate for the maximum time that is feasible for marking the Australian Physics Olympiad exam. Normally the marking occurs in a 48 hour period. This could be extended to approximately two weeks for the 1500 responses to 45 questions. This means that a sample of 200 responses to one question (as used in this experiment) should take approximately 4000 seconds to mark (20 seconds per response). This includes all steps, not just the preprocessing explored in this paper. 

The time could be scaled up by a factor of two or three by getting multiple computers and further extending the time period of marking. 
But either way, 4000 seconds is the order-of-magnitude time requirement and is shown as a blue vertical line on Figure \ref{fig:wall_time_accuracy_log}. 
We believe that typical physics instructors will have access to computational power which corresponds to this time,
as it is likely that they will have less students and more time, but less than 16GB of memory on their computer's graphics cards.

\subsection{Types of Translation Errors}
In this section we define error categories which would lead to an incorrect result when grading with an SMT or CAS. We then determine how frequently models made errors corresponding to each of these types.

\paragraph{Fictitious Attachment}
LLMs were far more likely to hallucinate equations when students referenced an attachment, even though the model was not shown any attachments. An example of fictitious attachment is this student response:  `Attachments: IMG\_4805.HEIC (905.5KB)', although this text was written, no attachment was given to the model. Interestingly, DeepSeek 671B and Llama405B reached a consensus that this response contained two equations: $v= u + wt$ and $S= ut + (1/2)wt^2$. This is consistent with the errors discovered by Filighera \cite{Filighera2024Cheating} that the presence of specific words can decrease LLM grading accuracy.\\

\paragraph{Undefined Symbols}
For the AlphaPhysics pipeline to work we need the response to be in a standardized format for the symbolic reasoner. Even trivial errors like incorrect capitalization will cause an SMT solver or CAS to fail. An example of an undefined symbol is as follows, a student typed `s/u - w = s/v + w'. The LLM interpreted this to be $s/u - w, s/v + w$, which is incorrect  because $s$ needs to be capitalised to $S$ in the standard format. In another example a student wrote `v = plusminus root(u + w $\hat{}$ 2)' which the model interpretted as two equations $v= \text{sqrt}(u + w^2))$ and $v= -\text{sqrt}(u + w^2)$. However `sqrt' is not a defined symbol and the models needed to instead write $v=(u + w^2)^{(1/2)}$ and $v= -(u + w^2)^{(1/2)}$.\\

\paragraph{Missed or Added Equation}
Some student responses included many equations. Sometimes the models missed one or more of these equations. There were also instances where extra equations were added by the model. An example of an extra equation is this student response `Group time: 2s/u seconds' which should be interpreted as $t= 2S/u$. Deepseek 671B and Llama405B instead arrived at a pair of equations: $t= 2S/u$ and $v=2S/u$, of which the second is a hallucination. \\

\paragraph{Incorrectly Interpreted Meaning}
There were some cases where students did not explicitly define variables and there was no well-known quantity that corresponded to the variable they chose.
For example a student may choose to use the generic, undefined `$x$'. There were two circumstances where the marking team was able to unanimously agree on the meaning of these variables from the context but the best models were not. The first  was this response: `s/(x + 0.7) + s/(x - 0.7) = s/u'. The model misinterpreted $x$ to be $t$ when the marking team unanimously decided that the student in fact meant $v$. The second example is as follows:
\begin{quote}
``$S / u = (0.5S / v + w) + (0.5S / v - w)$\\
$= (0.5S(v-u) + 0.5S(v+u))/(v^2 - u^2)$\\
$= Sv / (v^2 - u^2)$''
\end{quote}
Markers interpreted the first line of the student working to mean $S/u=0.5S/(v + w) + 0.5S/(v - w)$, but the model copied what the student literally wrote: $S/u= S/(2v) + w + S/(2v) - w$. 
The human markers were able to see that the second line logically follows from the first if there is a bracket which includes $w$ in the denominator (though it has changed to a $u$). Placement of this bracket also allows for dimensional consistency. 

\textbf{Summary of Error Types}
The models with the highest accuracy made a total of twenty-four errors in their translations. Each error was investigated manually for the consensus responses from Deepseek 671B and Llama 405B and classified into error types, see Table \ref{tab:error_types}. These models most commonly made errors when students made reference to an attachment that did not exist. 
In these cases the model would hallucinate equations. 

\begin{table}
  \centering
  \caption{Frequency of different error types in consensus responses for different model sizes. Models were paired with the other model of the same size for consensus, except for Llama3.1 405B and Deepseek 671B which were paired with each other. Evaluation of the 400B+ models was performed manually. For the other model sizes the frequency of each error determined automatically which means that the `Missed' and `Added' an equation are lower bounds. It possible for a single LLM response to contain more than one error type. }
  \label{tab:error_types}
  \setlength{\tabcolsep}{6pt} 
  \begin{tabular}{p{3.3cm} *{4}{c}}
    \hline
    \textbf{Type of Error} & \multicolumn{4}{c}{\textbf{Error in Consensus}} \\
        \cline{2-5}
         & 400B+ & 70B & 14B & 7B  \\
        \hline
        \multirow{2}{*}{\parbox{3.2cm}{Fictitious attachment / False Reference}} & 9 & 9 & 16 & 23 \\
        & & & & \\
        Used undefined Symbol & 6 & 11  & 17  & 52   \\
        Missed an equation & 4 & 7 & 9 & 41   \\
        Added an equation & 3 & 26 & 25 & 22   \\
        \multirow{2}{*}{\parbox{3.3cm}{Incorrectly interpreted meaning}} & 2 & NA & NA & NA  \\
        & & & & \\
        \hline
        Total Errors & 24 & 44 & 60 & 115  \\
    \hline
  \end{tabular}
\end{table}

We used an automated method of counting the number of equations and detecting the presence of undefined symbols to classify the error in responses of the models which contained 7, 14 and 70 billion parameters. 
These results shown in Table \ref{tab:error_types} are only a lower bound on the `missed an equation' and `added an equation' categories as it is possible that a model could have both missed and added an equation which would not impact the overall equation count.

There is an increase in most error types as the model size decreases. Adding an equation is the only exception where this remained roughly constant for models of 7-70 billion parameters.

\section{Discussion and Future Work}

In this section, the results of the experiments will be related to the practical challenges of marking Australian Physics Olympiad papers in the required time-frame. This is not meant to signal that the Australian Physics Olympiad is considering the use of automated marking tools for future exams. We will also discuss the wider implications of the findings.

\paragraph{Implications for Australian Physics Olympiads}
Figure \ref{fig:model_accuracy_log} shows that the highest accuracy score achieved by any model and condition was $88\%$. This is $8.5\%$ below the desired threshold of $96.5\%$ where the markers find the responses unambiguous.  Therefore further increases in accuracy would be required for a model to be sufficiently reliable to be included in the marking pipeline of the Australian Physics Olympiad Exam. One key area for future work would be to determine new model architectures or strategies to improve accuracy. One strategy which has been shown to work is to break the task into smaller components \cite{McGinness2024Highlighting}, which could be implemented by including only one line of the student response per model call instead of the entire response.

Figure \ref{fig:wall_time_accuracy_log} shows that some of the models are able to respond quickly enough that it would be feasible to use them grading. However, the thinking models which make use of excessive tokens run too slowly to be useful on the time-scale of the Australian Physics Olympiad exam. 
The results also indicate that the maximum feasible model size is approximately 14 billion parameters. 
Phi4 shows the most promise of these models, and decreasing the precision of the floating point numbers from 16 bits to 4 bits only marginally decreases its performance. 
Nearly all of the improvements to LLMs since December 2023 can be attributed to increasing the computational expense through either model size or number of tokens (Chain of Thought or thinking models) \cite{McGinness2025Imitate}. A new approach to improving model performance is needed to meet the speed and hardware requirements. 
Some recent innovations include diffusion-based language models \cite{Nie2025Large} or scaling test time compute using a recurrent block \cite{Geiping2025Scaling}.  \\

\paragraph{Wider Implications for LLM Physics Grading}
This study has shown that as of March 2025, the frontier open-source LLMs were able to extract equations from complex student responses at a level of approximately 88\% accuracy, which is still below the level of a team of human graders. 
Although this study did not consider the actual assigning of grades to responses, this level of accuracy is roughly in line with a previous study which found that the discrepancy between human-assigned and AI-assigned grades was characterized by $R^2=0.84$ \cite{Kortemeyer2023Toward}.

Instructors who are interested in automated grading solutions will (hopefully) be interested in an accurate solution which minimizes computational power and therefore both electricity usage and environmental impact. 
Figure \ref{fig:model_accuracy_log} indicates that increased model size correlates with overall performance. Interestingly, there is little improvement for thinking models beyond 70 billion parameters. Non-thinking models also reach this same performance plateau, just at a significantly larger model size. These results provide an early suggestion that there may be an optimum model size for computationally efficient, accurate physics grading. An area for future work would be to perform more experiments to confirm this trend and more precisely find the optimum value.\\

\paragraph{Computational cost} Physics instructors may be interested in quantifying the computational and electrical cost of LLM grading. In our previous studies \cite{McGinness2025Imitate} we have found that the well-known approximation \cite{Kaplan2020Scaling} for the number of Floating Point Operations (FLOPs) to generate an LLM response, given by Equation \ref{eqn:FLOPsApproximation}, is accurate within 10\%.
\begin{equation}
    \label{eqn:FLOPsApproximation}
    \text{FLOPs}=2Nn
\end{equation}

Where $N$ is the number of active model parameters and $n$ is the number of prompt tokens plus completion tokens. This means that the computational cost is linearly proportional to the number of active parameters of the model. Therefore improvement due to model size comes at a cost.

Equation \ref{eqn:FLOPsApproximation} implies that thinking models have an increased computational cost as they produce more tokens. Prompting strategies which have long context windows (like many-shot prompting) or produce large numbers of tokens (Chain of Thought) will also increase environmental impact. 

If an instructor was not concerned by privacy or transparency, but simply wanted to reduce environmental impact, a key consideration is whether LLMs should be run locally or on GPU servers.
We perform a basic analysis in Table \ref{tab:specs} based on official datasheets released by NIVIDA \cite{A100,H100,RTX4060} and estimates of global datacenter efficiency by the Uptime Institute \cite{Uptime2025Global}. 
These datasheets state the number of 16-bit (FP16) Terra-Floating Point Operations Per Second (T-FLOPS) that the GPU can perform at maximum capacity and the power consumption of the GPU in Watts. 
Once the overhead (for example energy used for cooling the system) is considered this can allows us to calculate the Terra-Floating Point OPerations (T-FLOPs) per Joule.

\begin{table}[h]
    \centering
    \caption{Comparison of local and server based GPU performance on sparse matrix multiplications. System power overhead was calculated for the RTX 4060Ti system using NVIDIA's estimated total required system power of 550 Watts \cite{RTX4060}. We used the Global PUE vales as system overhead for cloud based H100 and A100 GPUs. The FP16 T-FLOPS for the RTX 4060Ti is taken as the reported number of AI TOPS (Trillions of Operations Per Second)}
    \label{tab:specs}
    \begin{tabular}{lcccc}
        \hline
        GPU & Watts & Sparse FP16 & System Power & T-FLOPs/Joule \\
         & & T-FLOPS & Overhead Ratio & \\
        \hline
        H100 & 700 & 1979 & 1.5 & 1.88 \\
        A100 & 400 & 624 & 1.5 & 1.04 \\
        RTX & 160 & 353 & 3.43 & 0.64 \\
        \hline
    \end{tabular}
\end{table}

Table \ref{tab:specs} shows that running models locally increases the power consumption for the same task by a factor of approximately 3. Therefore if an instructor was interested in minimizing their environmental impact it would be more efficient (regardless of which model they choose) to run this model on a GPU server rather than a local machine. Although running models locally is less efficient, we note that running local LLMs makes the user aware of the amount of compute that they are using.

In this analysis we only considered electrical power consumption, not water consumption from the server clusters, which may be of more concern in some areas. 

A limitation of the study was that although the responses were marked by a team of graders, only one final mark was assigned for each student response. In addition the outputs of the LLM were not used to calculate a score for each response. These factors make it impossible for an inter-grader Quadratic Weighted Kappa (QWK) and LLM-human QWK values to be compared. 

In this work we only consider typed responses for algebraic expressions. Future work could consider other types of questions such as diagrams, plots, numerical responses and short answer questions. As students are allowed to submit hand-written responses, another key area for future work is the use of multi-modal LLMs to parse these.

\section{Conclusions}

We presented the two-step, neuro-symbolic AlphaPhysics pipeline for physics grading. We performed systematic experiments using open-source models in the first step of this pipeline: Large Language Model (LLM) extraction and translation of student  equations into a standardized format. The lists of equations extracted from each of 200 typed student responses to an Australian Physics Olympiad question were compared to the lists generated by the Olympiad marking team.

We tested neuro-symbolic frameworks where the LLM receives feedback from an automated reasoning engine. We found that these frameworks improved model accuracy only when LLMs were provided with specific feedback on how to improve their responses.

We quantified the accuracy of models of different sizes for this task.
Accuracy of LLMs increases with model size, but there is a plateau of approximately 85\% accuracy which occurs at approximately 70B parameters for thinking models. 

We found that a 14B parameter model (Phi4) was able to achieve this task with approximately 70\% accuracy. The computational cost of using Phi4 to complete this task was small enough that a physics instructor could run it on their own computer. Although running models locally may alleviate privacy concerns, this is likely to increase the environmental impact.

Interestingly we found that if students made reference to an attachment that this increased the chance that LLMs of all sizes would hallucinate equations. This was the most common type of error for large models. Smaller models were more likely to use undefined symbols, miss an equation and hallucinate equations. 

\begin{acknowledgments}
We would like to thank Australian Science Innovations for providing access to student responses to the 2023 Australian Physics Olympiad Exam.

This project has been funded by the CSIRO and ANU. One of the authors of this paper is the director of the Australian Physics Olympiad program. 

The ethical aspects of this research have been approved by the ANU Human Research Ethics Committee (Protocol 2023/1362).
\end{acknowledgments}

\appendix

\section{Full Exam Question: Amusing Airport Adventures}
\label{app:question}
\small{
Some students are wandering around an airport waiting for their flight. 
This question deals with kinematics only, so there is no need to consider resistive forces such as drag.

A group of students are walking at a speed of 1.0m/s. They walk alongside a travellator which is moving at a speed of 0.70m/s. 

\begin{itemize}
    \item[1.] One of the students, Mali, decides to walk on the travellator which is moving in the same direction as the group of students. How fast must Mali walk to move at the same speed as the rest of the group? (1 mark)
\end{itemize}
Note when we refer to Mali's walking speed we mean with respect to the ground/travellator beneath them. 
\begin{itemize}
    \item[2.] The group stops to read a sign. Mali is still on the travellator. How fast must Mali walk now to stay with the group? (1 mark)
    \item[3.] The group moves at 1.0m/s towards the left. The travellator moves at 0.70m/s towards the left. Mali walks at a constant speed v relative to the ground/travellator beneath them. Mali wants to take path L and arrive at point A at the same time as the rest of the group. At what speed should Mali move? Ignore any distance traveled in the y-direction for this problem. (4 marks)
\end{itemize}
}

\begin{figure}[h]
    \centering
    \includegraphics[width=0.42\textwidth]{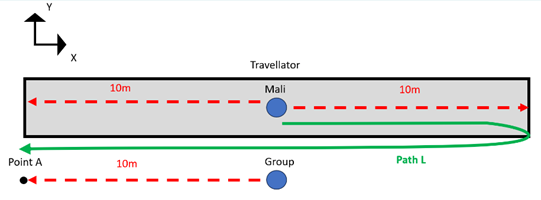}
    \label{fig:2023Q3}
\end{figure}

\small{
\begin{itemize}
    \item[4.] You may ignore the y-direction for this question; it exists simply for the clarity of the diagram. Imagine Mali and the group start in line with point A. They walk to the end of the travellator (point B) and then return. Mali completes the entire journey on the travellator. The group does the entire journey on the ground.
    \item[] Assume the group walking speed is u. You may not assume that u = 1.0m/s. Assume that the travellator speed is w. You may not assume that w=0.70m/s. Assume the travellator has length S.
    \item[] Mali walks at a constant speed, v (relative to the ground beneath them), for the entire journey. Determine speed v such that Mali and the group return at the same time. Give your answer as an algebraic expression.
    \item[] Explore the limiting behaviours of your equation. Explain what happens to v when u is very large compared to w, and what happens when w is very large compared to u. What happens to v when S is increased? Does the behaviour of your equation make sense in these scenarios? Why? (7 marks)
\end{itemize}}

\begin{figure}[h]
    \centering
    \includegraphics[width=0.42\textwidth]{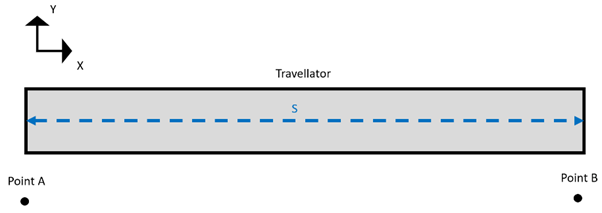}
    \label{fig:2023Q4}
\end{figure}

\section{Selected Models}
\label{app:models}

Table \ref{tab:models} classifies the Large Language Models used in the experiments into thinking or non-thinking and by number of parameters.

\begin{table}[hbp]
    \centering
    \caption{List of Open-Source models used in the experiments, classified into either thinking or non-thinking models. Larger models are slower and more expensive to run, therefore many companies will create smaller `distilled' models. Distillation is a technique where a small model is trained to produce similar outputs to a larger one \cite{Sanh2019DistilBert}.}
    \label{tab:models}
    \scalebox{0.75}{
    \begin{tabular}{p{2.1cm}@{\hspace{5mm}}p{4cm}@{\hspace{7mm}}p{4cm}}
        \toprule
        \textbf{Model Size (Number of Parameters)} & \textbf{Non-thinking Model} & \textbf{Thinking Model} \\
        \midrule
        1-2B & Llama3.2 1B \cite{Llama3.2} & DeepSeek 1.5B \cite{DeepSeekDistilled} \\
        & \small{A small model released by MetaAI in 2024} & \small{A Qwen distilled version of Deepseek 671B} \\
        \addlinespace
        7B & Mathstral \cite{Mathstral} & DeepSeek 7B \cite{DeepSeekDistilled} \\
        & \small{A small model released by Mistral AI in 2024} & \small{A Qwen distilled version of Deepseek 671B} \\
        \addlinespace
        14B & Phi4 \cite{Phi4} & DeepSeek 14B \cite{DeepSeekDistilled} \\
        & \small{A model released by Microsoft in 2024} & \small{A Qwen distilled version of Deepseek 671B} \\
        \addlinespace
        70B & Llama3.3 70B \cite{Llama3.3} & DeepSeek 70B \cite{DeepSeekDistilled} \\
        & \small{A model released by MetaAI in 2024} & \small{A Llama distilled version of Deepseek 671B} \\
        \addlinespace
        400B+ & Llama3.1 405B \cite{Llama3.1} & DeepSeek 671B \cite{DeepSeek} \\
        & \small{A large model released by MetaAI in 2024} & \small{A large model released by DeepSeek AI in January 2025} \\
        \bottomrule
    \end{tabular}}
\end{table}



\section{Additional Algorithms}
\label{app:algorithms}

This section contains the helper functions for Algorithms \ref{alg:llm-modulo} and \ref{alg:consensus}. Algorithm \ref{alg:attempt-parse} (Attempt\_parse) returns True and None if no formulas in a list contain syntax errors. If there are error messages it returns false and the error messages.

\begin{algorithm}
\caption{Attempt\_parse: Checks and Parses LLM Equations}
\label{alg:attempt-parse}
\KwIn{$equation\_list$}
\KwOut{$correct\_syntax$ (boolean), $error\_messages$ (if applicable)}

\Try{
    Parse $equation\_list$ through system\;
    \Return{\textbf{True}, None}\;
}
\Catch{\text{ParsingError}\;
        \Return{\textbf{False}, error\_messages}\;
    }
\end{algorithm}

Algorithm \ref{alg:check-equivalence} checks whether two lists of student equations are equivalent using the process described for equivalence in \ref{sec:metrics}.

\begin{algorithm}
\caption{Check\_Equivalence: of equation Lists}
\label{alg:check-equivalence}

\KwIn{$student\_equations$, $llm\_equations$}
\KwOut{\textbf{True} if lists are equivalent; otherwise \textbf{False}}
\SetKwFunction{FEquiv}{Z3Entailment}
\SetKwProg{Fn}{Function}{:}{end}

\ForEach{$eq_{student}$ in $student\_equations$}{
    $found \gets False$\;
    \ForEach{$eq_{llm}$ in $llm\_equations$}{
        \If{\FEquiv{$eq_{student}$, $eq_{llm}$}}{
            $found \gets True$\;
            \textbf{break}\;
        }
    }
    \If{$found = False$}{
        \Return{\textbf{False}}\;
    }
}

\ForEach{$eq_{llm}$ in $llm\_equations$}{
    $found \gets False$\;
    \ForEach{$eq_{student}$ in $student\_equations$}{
        \If{\FEquiv{$eq_{llm}$, $eq_{student}$}}{
            $found \gets True$\;
            \textbf{break}\;
        }
    }
    \If{$found = False$}{
        \Return{\textbf{False}}\;
    }
}

\Return{\textbf{True}}\;
\end{algorithm}

Algorithm \ref{alg:z3-equivalent} describes the \texttt{Z3Entailment} function used in Algorithm \ref{alg:check-equivalence}. This function calls an SMT solver to check if two equations can be shown to be unsatisfiable.

\begin{algorithm}
\caption{Z3Entailment: Check Equation Entailment using Z3}
\label{alg:z3-equivalent}
\KwIn{$eq_1$, $eq_2$}
\KwOut{\textbf{True} if $eq_1$ entails $eq_2$; otherwise \textbf{False}}

    \Try{
        $eq1\_z3 \gets ParseToZ3(eq_{1})$\;
        $eq2\_z3 \gets ParseToZ3(eq_{2})$\;
    }
    \Catch{\text{ParsingError}\;
        \Return{\textbf{False}}\;
    }

    $prog \gets InitializeZ3Program()$ \;
    
    \tcc{Refutational formulation of checking whether eq1 entails eq2:}
    
    Add constraint $eq1\_z3$ to $prog$\;
    Add constraint $\neg(eq2\_z3)$ to $prog$\;

    $result \gets ExecuteZ3(prog)$\;

    \eIf{$result = UNSAT$}{
        \Return{\textbf{True}}\;
    }{
        \Return{\textbf{False}}\;
    }
\end{algorithm}

\FloatBarrier

\section{Extra Data Tables}
\label{app:results}

Once the initial experiments were conducted, additional targeted experiments were performed for models with strong performance just above the reasonable time requirement. 
The parameters of LLMs are stored as floating-point numbers which are normally represented by 16 bits. 
However it is possible to round the numbers to less significant figures allowing them to be stored in 8 bits or 4 bits, referred to as 8-bit or 4-bit quantisation respectively. 
This has the advantage of using less memory and requring less computation per floating point operation (FLOP). 
We performed experiments with these lower quantisations for the chosen models and
the results are shown in Table \ref{tab:raw-data2}. Llama 3.3 70B and Deepseek-r1 70B were able to be run on a local machine using 4-bit floating point numbers (quantisation).

\begin{table}[htbp]
    \centering
    \caption{Raw data for the additional experiments.}
    \label{tab:raw-data2}
    \begin{tabular}{l|*{4}{c}}
        \hline
        \textbf{Model} & 
        \multicolumn{1}{c}{Llama 3.3} & 
        \multicolumn{1}{c}{Deepseek} & 
        \multicolumn{1}{c}{Phi4} & 
        \multicolumn{1}{c}{Phi4} \\
        \hline
        Model Size (B) & 70 & 70 & 14 & 14 \\
        Quantisation (bit) & 4 & 4 & 8 & 4 \\
        \hline
        \multicolumn{5}{c}{\textbf{Direct}} \\
        \hline
        Time (s) & 6,763 & 89,566 & 624 & 314 \\
        Tokens & 5,268 & 89,871 & 5,471 & 5,497 \\
        Correct Answers & 134 & 133 & 142 & 141 \\
        \hline
        \multicolumn{5}{c}{\textbf{LLM-Modulo SymPy}} \\
        \hline
        Time (s) & 6,978 & 90,625 & 651 & 337 \\
        Tokens & 5,461 & 91,003 & 5,974 & 5,732 \\
        Correct Answers & 134 & 133 & 142 & 141 \\
        \hline
        After Consensus & 144 & 148 & 141 & 141 \\
        \hline
        \multicolumn{5}{c}{\textbf{LLM-Modulo Z3}} \\
        \hline
        Time (s) & 7,183 & 126,033 & 796 & 377 \\
        Tokens & 5,537 & 99,298 & 7,104 & 6,785 \\
        Correct Answers & 141 & 146 & 147 & 144 \\
        \hline
    \end{tabular}%
\end{table}

\section{Prompts}
This section contains the exact prompts used to call the Large Language Models for each of the tasks. The first prompt is the prompt that was used for the direct method, and therefore also the first step in the LLM-Modulo SymPy, LLM-Modulo Z3 and Consensus methods:

{\scriptsize
\begin{verbatim}
You are an expert in interpreting student mathematical responses 
to physics questions. Please read this student answer: 
<<<{student_answer}>>>

When writing equations, use SymPy syntax following these 
guidelines exactly:
   - Use Eq(left, right) syntax, e.g., "Eq(v, (2*u) + w)"
   - If student writes an expression without 
    "{EXPECTED_VARIABLES[0]}=", assume that this is intended to 
    be the right hand side of the equation and the left side 
    would be {EXPECTED_VARIABLES[0]}
   - Separate multiple equations with commas
   - Only use these variables: {EXPECTED_VARIABLES}. 
   Capitalisation of the variables is very important, please 
   adjust the student's variables to match the capatilisation 
   of the expected variables given here.
   - Correct notation (for example v0 → v_0 or x^2 → x**2) but
   do not correct math errors
   - Use brackets for proper order of operations
   - Use "**(1/2)" for square root
   - If there are multiple equations, separate them by commas
   but include all of them within "[" and "]".
   - Use sin(x) and cos(x) for trigonometric functions, as in 
   SymPy syntax. For example, "sin(x)" should remain sin(x) 
   and "cos^2(theta)" should be written as (cos(theta)**2).
   - Note that special symbols may not have formatted 
   correctly. Therefore if you see "?" there is a high chance
   that the student meant "theta", "phi" or another Greek 
   letter.
   Examples:
   - Student writes "E=2u+w": Output should be 
   "Eq(E_0, (2*u) + w)"
   - Student writes "E0=u^2/w": Output should be 
   "Eq(E_0, (u**2)/w)" 
   - Student writes "sin x": Output should be 
   "Eq(E_0, sin(x))"
   - Student writes "v=sin(x) + cos^2(theta)" Output should be 
   "Eq(v, sin(x) + (cos(theta)**2))"

The based on these guidlines analyse the provided student 
answer and complete the following task exactly:
Write "List of Equations: [". Then write all equations 
that are contained in the student text using SymPy format 
following the guidelines above. Separate each equation 
with a comma (,).  Then write "]".  If the student has 
not written an equation using symbols (only describing 
in words) then leave the list blank "[]".

When completing the task, if the student writes any 
variables that are not in this list: {EXPECTED_VARIABLES} 
use your expert judgement to interpret and rewrite what 
they meant in terms of these variables.
Use the exact marker text and provide no additional 
text or justification.

Student answer: <<<{student_answer}>>>
Your response to the task:
\end{verbatim}
}

The following repair prompt was used in the LLM-Modulo methods:
{\scriptsize
\begin{verbatim}
You are an expert in interpreting student mathematical responses
and converting them into valid SymPy syntax. 
Your overall goal is to extract all equations that students have
explicitly written in their response to a physics question. 
Note that equations described in words do not count, only record 
equations that students have written in SymPy syntax.
The student may have made syntactic errors, please use your 
expertise to interpret what the student meant and write record 
the syntactically correct version.
Ensure that only the expected variables are used, they are listed
here: {EXPECTED_VARIABLES}

The student's answer is as follows: <<<{student_answer}>>>
Previously you recorded this equation: {original_response}
However it could not be parsed, giving this error: {error_msg}

Your task is to repair your previous response which had a 
parsing error. Learn from the error messages, try changing symbols
or operators to fit the required syntax and don't repeat the 
same error again. 

Guidelines for your corrected response:
1. Begin with "List of Equations: [" and end with "]".
2. All equations must be in SymPy format using Eq(left, right), 
e.g., "Eq(v, u + w)" for "v = u + w"
3. If student writes expression without "{EXPECTED_VARIABLES[0]}=",
assume that this is intended to be the right hand side of the 
equation and the left side would be {EXPECTED_VARIABLES[0]}
4. If multiple equations, separate them with commas. Do not use \n.
5. Only use these variables: {EXPECTED_VARIABLES}. Capitalisation 
of the variables is very important, please adjust the student's 
variables to match the capatilisation of the expected variables 
given here.
6. Correct notation errors (capitalization, brackets) but not 
mathematical errors
7. Use "**(1/2)" for square root
8. If there are multiple equations, separate them by commas but 
include all of them within "[" and "]". Do not use square brackets 
for any other purpose.
9. Use sin(x) and cos(x) for trigonometric functions, as in SymPy 
syntax. For example, "sin(x)" should remain sin(x) and 
"cos^2(theta)" should be written as (cos(theta)**2).
10. Note that special symbols may not have formatted correctly. 
Therefore if you see "?" there is a high chance that the student 
meant "theta", "phi" or another Greek letter.
11. Example formats (pay careful attention to brackets and order 
of operations):
   - Student writes "E=2u+w": Output should be "Eq(E_0, (2*u) + w)"
   - Student writes "E0=u^2/w": Output should be "Eq(E_0, (u**2)/w)"
   - Student writes "sin x": Output should be "Eq(E_0, sin(x))"
   - Student writes "v=sin(x) + cos^2(theta)" Output should be 
   "Eq(v, sin(x) + (cos(theta)**2))"

Provide only the "List of Equations: [...]" response, with no 
explanation or justification.
\end{verbatim}
}

This prompt was used in the consensus condition:
{\scriptsize
\begin{verbatim}
You are an expert in interpreting student mathematical responses
and converting them into valid SymPy syntax.
        
Two different interpretations were given for a student's answer, 
and we need your expert analysis to determine the most accurate 
interpretation.

Original student answer: <<<{student_answer}>>>
{model1_id} interpretation: {response1}
{model2_id} interpretation: {response2}

Guidelines for your consensus response:
1. Begin with "List of Equations: [". 
2. Then write the equation(s) in the student response. Separate 
them by commas, do not use \n.
3. Finally end your response with "]".

All equations must be in SymPy format using Eq(left, right).
- If student writes expression without 
"{EXPECTED_VARIABLES[0]}=", assume that this is intended to be 
the right hand side of the equation and the left side would be 
{EXPECTED_VARIABLES[0]}
- Separate multiple equations with commas
- Only use these variables: {EXPECTED_VARIABLES}. Capitalisation 
of the variables is very important, please adjust the student's 
variables to match the capatilisation of the expected variables 
given here.
- Correct notation (v0 → v_0, x^2 → x**2) but not math errors
- Use brackets for proper order of operations
- Use "**(1/2)" for square root
- If there are multiple equations, separate them by commas 
but include all of them within "[" and "]".
- Use sin(x) and cos(x) for trigonometric functions, as in 
SymPy syntax. For example, "sin(x)" should remain sin(x) 
and "cos^2(theta)" should be written as (cos(theta)**2).
- Note that special symbols may not have formatted correctly. 
Therefore if you see "?" there is a high chance that the 
student meant "theta", "phi" or another Greek letter.
Examples:
- Student writes "E=2u+w": Output should be 
"Eq(E_0, (2*u) + w)"
- Student writes "E0=u^2/w": Output should be 
"Eq(E_0, (u**2)/w)" 
- Student writes "sin x": Output should be 
"Eq(E_0, sin(x))"
- Student writes "v=sin(x) + cos^2(theta)" Output should be 
"Eq(v, sin(x) + (cos(theta)**2))"

Analyze both interpretations and the original student answer, 
then provide your expert consensus in the exact format 
specified above. 
Provide only the "List of Equations: [...]" response, with no 
explanation or justification. 
\end{verbatim}
}

\FloatBarrier

\bibliography{PhDThesisReferences}

\end{document}